\documentclass[3p,times,nopreprintline]{elsarticle}

\let\today\relax
\makeatletter
\def\ps@pprintTitle{%
    \let\@oddhead\@empty
    \let\@evenhead\@empty
    \def\@oddfoot{\footnotesize\itshape
         {Submitted preprint} \hfill\today}%
    \let\@evenfoot\@oddfoot
    }
\makeatother

\usepackage{ecrc}
\usepackage[bookmarks=false]{hyperref}
    \hypersetup{colorlinks,
      linkcolor=blue,
      citecolor=blue,
      urlcolor=blue}
\usepackage{subcaption}

\volume{00}

\firstpage{1}

\runauth{J. Žižka}

\usepackage{amssymb}
\usepackage[figuresright]{rotating}
\usepackage[T1]{fontenc}

\begin{document}
\begin{frontmatter}

\title{Statistical complexity of software systems represented as multi-layer networks}

\author[a]{Jan Žižka} 

\address[a]{Masaryk University, Faculty of Informatics, Botanická 68a, Brno 60200, Czech Republic}

\begin{abstract}
Software systems are expansive, exhibiting behaviors characteristic
of complex systems, such as self-organization and emergence. 
These systems, highlighted by advancements in Large Language Models (LLMs)
and other AI applications developed by entities like DeepMind and OpenAI
showcase remarkable properties. Despite these advancements, there is a notable
absence of effective tools for empirically measuring software system complexity,
hindering our ability to compare these systems or assess the impact of
modifications on their properties. Addressing this gap, we propose
the adoption of statistical complexity—a metric already applied in fields
such as physics, biology, and economics—as an empirical measure for evaluating
the complexity of software systems. Our approach involves calculating the statistical
complexity of software systems modeled as multi-layer networks
validated by simulations and theoretical comparisons. This measure offers
insights into the organizational structure of software systems,
exhibits promising consistency with theoretical expectations,
and paves the way for leveraging statistical complexity as a tool
to deepen our understanding of complex software systems and
into their plausible and unplausible emergent behaviors.
\end{abstract}

\begin{keyword}
Statistical Complexity; Software Systems; Complex System Theory; Complex Software Systems; Multi-layer Networks
\end{keyword}

\cortext[cor1]{Corresponding author.Email: jzi@mail.muni.cz}

\end{frontmatter}


\section{Introduction}

Software systems are continuously growing in size and complexity,
becoming increasingly large and complicated~\cite{koch2007software},
with a super-linear growth rate. Some of these systems are starting to
exhibit properties and features of complex systems, such as self-organization
and emergence, which are bringing surprising advancements to our society,
as seen in systems developed by DeepMind~\cite{deepmind}, and OpenAI~\cite{openai}.
However, we lack tools for designing and
building large systems with valuable complex behaviors.
Complex software systems may be
defined by a set of properties~\cite{zizka2023towards}, but these do not allow for calculable empirical values.
Understanding software systems requires a combination of tools and viewpoints.
The motivation for our work is to deepen our knowledge of complex software systems,
providing insight into plausible emergent behaviors, such as those bringing added
value to software system use beyond their original design, as well as unplausible
emergent behaviors, such as emergent errors or increased security attack surface
or indicators that the software system is complicated rather than complex, affecting
its maintenance and development costs.

In this paper, we propose using statistical complexity measures~\cite{Crutchfield1989, LopezRui1995}
as a methodological tool to calculate the complexity of software systems.
This measure may provide a means to compare different software systems
based on their complexity and measure the effects of different
properties on software system complexity.
Statistical complexity appears to describe system complexity well~\cite{LopezRui1995}
as it allows for the disqualification of purely ordered and fully
chaotic systems. It measures the information
in the system and the emergence of structure or patterns.

We hypothesize that statistical complexity can be calculated for software
systems represented as multi-layer networks, where the state of the system
is represented by the measure of inter-component communication and its internal
state. We expect a completely ordered system where only one state out
of all possible system states is realized will have zero statistical complexity.
Conversely, a software system where the communication and component state is random,
chaotic, or disordered and lacks internal organization will have
low statistical complexity. We expect that a software system with an internal structure
represented as a multi-layer network with ordered relationships among the layers
will exhibit higher values of statistical complexity. 
We surmise that a system organization
and information content exists that provides the highest
statistical complexity. We suggest that software systems of a similar type,
with the system state represented by the same qualitative measures, can be
compared using normalized statistical complexity, providing information on
which of the systems is more complex from the viewpoint of this measure,
indicating where the system’s organization and amount of information are the highest.
We will validate the hypothesis by running simulations of software systems modeled
as multi-layer networks with varying parameters, such as the number of components,
layers, and simulation time. The results will be compared to the
expected hypothesis.

The contributions of this paper are:
Formulating equations to calculate the statistical complexity of
a software system.
Specifying the design of a software system model as a multi-layer
network and the selection of a simulation framework.
Providing validation of the hypothesis that statistical complexity
calculated for a software system exhibits the expected behavior
based on simulation.
Demonstrating that statistical complexity can be used to compare
software systems and that it reflects the software system’s organization
and information content.

The paper is structured as follows. First, we discuss software
system complexity measures in Section~\ref{sec:software-complexity}, where we
introduce statistical complexity. In Section~\ref{sec:validation}, we describe
the design of the software system model, and we share
the discussion on simulation execution and obtained results.
Observed limitations of the approach are presented
in Section~\ref{sec:limitations}, and conclusions are shared in Section~\ref{sec:conclusions}.

\section{Software System Complexity Measure}
\label{sec:software-complexity}

\subsection{System Complexity Calculation Methods}
\label{subsec:methods}

The literature proposes multiple methods for calculating a value corresponding
to the complexity of a system~\cite{Ladyman2013,LopezRui1995,tabilo2023brief},
which might be applied to software systems.
In this section, we briefly overview some of the methods and discuss, in more detail, those accepted as suitable measures of complexity.
The methods are summarized in Table~\ref{tab:methods}.

\textbf{Kolmogorov Complexity:}
One of the primary system complexity measures is the algorithmic complexity,
also known as Kolmogorov complexity~\cite{kolmogorov1965three}.
The value of Kolmogorov complexity of an object, in our case, a system,
defines the shortest program that can produce the object as output
when run on a universal Turing machine~\cite{turing1936computable}.
The Kolmogorov’s complexity is generally uncomputable,
as already noted by Kolmogorov. Although Kolmogorov's complexity may
be estimated~\cite{evans2002kolmogorov}, but even that is computationally difficult.
The Kolmogorov's complexity is context-dependent~\cite{edmonds1999syntactic},
complicating its general usability. Using Kolmogorov complexity
to measure software system complexity might be challenging.

\textbf{Effective Complexity:}
Another proposed measure of complexity is effective complexity~\cite{gell2002complexity}.
Defined by the length of a concise description
of an object’s regularities, effective complexity measures the amount of
non-random information in the system. Like Kolmogorov’s complexity,
effective complexity is context-dependent and lacks a universally accepted
calculation method, making it impractical to measure the complexity
of software systems.

\textbf{Shannon Entropy:}
Originating from thermodynamics and information
theory~\cite{shannon1948mathematical}, quantifies the uncertainty
in a system’s state, with ordered systems having low Shannon entropy
and disordered or chaotic systems having high Shannon entropy.
The Shannon's entropy does not capture the system’s organization or structure,
thus failing to distinguish between complex and chaotic systems.

\textbf{Statistical Complexity:}
Combines information
quantity and system organization measures~\cite{Crutchfield1989}. A fully ordered system has zero statistical
complexity. As the organization of a system increases along with its complexity,
the statistical complexity increases, while a fully chaotic system has again
low statistical complexity.
Statistical complexity will be discussed in detail in the following sections.

\textbf{Network Theory Based Complexity:}
The work by Gao \& Li~\cite{gao2009complex} measures software system complexity
using complex network theory, introducing a system as a network, and
deriving complexity calculations based on the distance to “average”
values representing the system. This method, conceptually similar
to statistical complexity, focuses primarily on structural complexity
without considering information or entropy.

\textbf{Graph Theory Based Complexity:}
Dali~\cite{dali2014complexity} utilizes network graph theory to measure
a system’s structural complexity based on various metrics.
The proposed method utilizes network graph theory
to determine the structural complexity of a system based on \emph{$d$}
the average shortest distance, \emph{$p_k$} the degree of distribution,
\emph{$C$} the average clustering coefficient, \emph{$B$} the betweenness,
and the correlation \emph{$R(k_i,k_j)$}, \emph{$R(k_i,C_i)$} determining
collaboration relation and modularization degree.
This method does not account for information content,
potentially misrepresenting ordered systems as complex.
Publications~\cite{dali2014complexity, ma2010hybrid} provide evidence that software systems exhibit
small-world and scale-free network properties, based on analysis
of software systems and calculated characteristic parameters.
With this proposed measure, an ultimately ordered system would show
high complexity measures.
Hanyan~\cite{hanyan2017software} extends work of Dali~\cite{dali2014complexity} and
Ma~et~al.~\cite{ma2010hybrid} with modeling a software system with three-dimensionality
layers, the structure level, the function level, and the code level.

\textbf{Generalized Framework:}
Efatmaneshnik~\cite{efatmaneshnik2016general} proposes a framework combining subjective
and objective measures of system complexity, demonstrating that statistical
measures fit within this framework.
The author introduces a subjective simplicity and proposes to calculate a subjective
complexity as a distance from simple measured as, following the Kullback-Leibler distance, topological difference.
The subjectivity depends on the selection of
a suitable reference model, and the author demonstrates that statistical measures of complexity
fit into the general framework of measuring complexity. Other measures
shown to follow the proposed general framework are those offered by Shiner~\cite{shiner1999simple};
however, this measure was criticized by Crutchfield~\cite{crutchfield2000comment}
by "... pointing out that it is over universal, in the sense that it has the same dependence
on disorder for structurally distinct systems."
Efatmaneshnik~\cite{efatmaneshnik2016general} also proposes that the cyclomatic complexity
measure follows the proposed general framework. The authors then suggest that the
engineered systems opt to be described by graph theory as the most
appropriate tool. 
The proposed generalized framework can also utilize the self-dissimilarity defined by Wolpert~\cite{wolpert2007using}
described by the Maximum Common Subgraph (MCS). Extracting  MCS is not
straightforward, and finding MCS is an NP-hard problem~\cite{bunke1998graph}.
The size of a system can be measured as graph entropy~\cite{mowshowitz2012entropy}
or by counting the number of graph elements (nodes and links).
Determining the simplicity is itself subjective and therefore unreliable.
The authors conclude that complexity measures tend to be context-dependent; 
The useful complexity measure for engineered systems has two components,
objective and subjective, which “allows to study system complexity from
the perspective of multiple stakeholders.”

\begin{table}[h]
    \caption{comparison of system complexity calculation methods}
    \label{tab:methods}
    \begin{tabular*}{\hsize}{@{\extracolsep{\fill}}cccccc@{}}
        \toprule
        Method & Calculable & Agreed methods & Information & Structure & Context independent \\
        \colrule
        Kolmogorov complexity & - & - & + & + & - \\
        Effective complexity & + & - & - & + & - \\
        Shannon entropy    & + & + & + & - & +  \\
        Statistical complexity   & + & + & + & + & -/+  \\
        Network theory based complexity   & + & + & - & + & -/+  \\
        Graph theory based complexity  & +/- & + & - & + & -/+  \\
        \botrule
    \end{tabular*}
\end{table}


Several methods proposed by literature for
calculation of system complexity were discussed in this section. A suitable method for
calculating the complexity of software systems requires the following
properties:
i) Must be calculable,
ii) must provide low complexity values for fully ordered as well as for disordered systems,
and if possible, iii) should be context-independent.

Based on the discussed properties, statistical complexity emerges as a suitable
measure of system complexity for software systems due to its ability to account
for ordered and disordered systems while preferably being context-independent.
Graph-based metrics show promise; however, their emphasis on structural aspects rather than informational content, combined with the challenges in calculating them, make them less ideal as a universal measure of complexity.
We focus on utilizing statistical complexity as a
measure of software system complexity.

\subsection{Statistical Complexity}
\label{subsec:statistical-complexity}

The Statistical Complexity Measure (SCM)~\cite{Crutchfield1989, LopezRui1995} is a concept used to quantify
the complexity of a system, combining elements of randomness (entropy)
and structure. The most common formulation of SCM is based on the concept
of Shannon entropy and the Jensen-Shannon divergence. The formula for SCM is typically given as follows:

\begin{equation}
SCM = H \cdot Q
\end{equation}

Where:
\( SCM \) is the statistical complexity,
\( H \) is the Shannon entropy, which measures the randomness or unpredictability in the system, and
\( Q \) is the disequilibrium, which is often quantified using a measure such as the
Jensen-Shannon divergence. It measures how different the actual system state probability distribution
is from a uniform distribution (equidistribution).
The Shannon entropy \( H \) is calculated as:

\begin{equation}
H = -\sum_{i} p_i \log(p_i)
\end{equation}

where \( p_i \) is the probability of the \( i \)-th state.
The disequilibrium \( Q \), using Jensen-Shannon divergence, is calculated as:

\begin{equation}
Q = JSD(P || R)
\end{equation}

Where \( P \) is the actual probability distribution of the system, \( R \)
is the reference probability distribution (often taken as the uniform distribution),
and \( JSD \) is the Jensen-Shannon divergence.
The Jensen-Shannon divergence itself is defined as:

\begin{equation}
JSD(P || R) = \frac{1}{2} D(P || M) + \frac{1}{2} D(R || M)
\end{equation}

where \( M \) is the average of \( P \) and \( R \), and \( D \) is the Kullback-Leibler divergence, given by:

\begin{equation}
D(P || M) = \sum_{i} p_i \log\left(\frac{p_i}{m_i}\right)
\end{equation}

and similarly

\begin{equation}
D(R || M) = \sum_{i} r_i \log\left(\frac{r_i}{m_i}\right)
\end{equation}

\begin{figure}[h]\vspace*{10pt}
    \centerline{\includegraphics[width=0.35\textwidth]{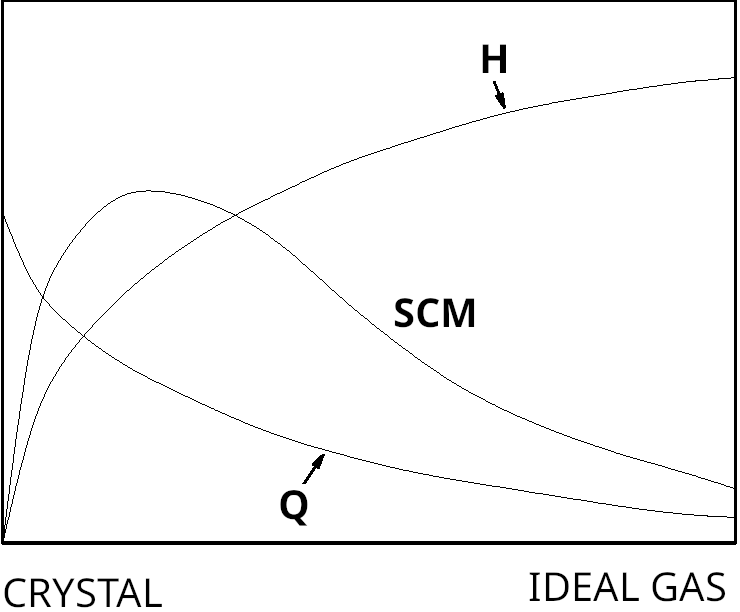}}
    \caption{Sketch of the intuitive notion of the magnitudes of “information” (H)
    and “disequilibrium” (Q) for the physical systems and the behavior intuitively
    required for the magnitude of “complexity.” The quantity SCM = H·Q is proposed
    to measure such a magnitude.~\cite{LopezRui1995}}
    \label{fig:scm-scatch}
\end{figure}

SCM combines the concepts of entropy and divergence
to provide a measure of complexity that accounts for both randomness
and pattern or structure in the system. It is used in complexity science,
information theory, and statistical physics.
Figure~\ref{fig:scm-scatch}~\cite{LopezRui1995} demonstrates SCM with respect to the system organization
for physics systems ranging from a crystal to an ideal gas. It shows how the entropy
and the disequilibrium contribute to statistical complexity.
It also demonstrates the fact that a completely disordered system, as well
as a completely ordered system has low statistical complexity.

The statistical complexity as proposed by Lòpez-Ruiz,~Mancini,~and~Calbet~\cite{LopezRui1995}
has to have a quantified clear interpretation as pointed out by Feldman~\cite{Feldman1998}.
This implies that the use of the SCM relies on the knowledge of the expected
structure of the system. In software systems, the engineered systems,
such knowledge lies within the software system design and implementation,
allowing for determining and quantifying the statistical complexity results.


Both the magnitude of the Shannon entropy $H$ and disequilibrium $Q$ depend
on the number of states of the system. The values must be normalized to compare
the statistical complexity of systems with different numbers of components
and different numbers of states.
The entropy $H$ is directly correlated to the number of states; therefore
it can be normalized into the range 0 to 1 with:

\begin{equation}
    \overline{H} = \frac{H}{n \cdot log(n-1)}
\end{equation}

Note that we do construct the system state probability vector by
concatenating component state probabilities
(see Equation~\ref{equ:system-probabilities}), therefore we must normalize $H$ with
$n \cdot log(n-1)$ where $n$ in number of components in
the system.
The disequilibrium $Q$ represented with Jensen-Shannon divergence
does not have a direct correlation to probabilities of the system
state vector. 
%
%
For complex systems, $Q_{max}$ might be
estimated empirically by simulating or enumerating various possible
distributions and calculating which yields the maximum divergence from $R$.
We expect that an \emph{ordered} system configuration has
the maximum $Q$ compared to any other system configuration with
same number of components. We define $Q_{max} = Q_{Ordered}$
and the normalized $\overline{Q}$ can be calculated as:

\begin{equation}
    \overline{Q} = \frac{Q}{Q_{max}}
\end{equation}


To determine the system state probabilities, the number
of messages sent among components are collected. The state vector
is defined as:

\begin{equation}
    {v_i} = [s_1, s_2, \ldots, s_{n-1}]
\end{equation}

Where: \( {v_i} \) is the state vector of the \(i\)-th component,
\( s_j \) represents the count of messages sent to the \(j\)-th
component in the system, for \(j = 1, 2, \ldots, n-1\).
\(n\) is the number of components in the system.
To calculate the probabilities of each state as defined by the number of messages
between components in the system, we can normalize the vector $v_i$ by calculating the total number
of messages $S$:

\begin{equation}
    S = \sum_{i=1}^{n-1} s_i
\end{equation}

Utilizing $S$ we define the vector of probabilities \( {p} = [p_1, p_2, \ldots, p_n] \),
where each element \( p_j \) is calculated as the fraction of the \( j \)-th element of \( {v_i} \)
divided by \( S \). Therefore, each \( p_j \) can be defined as:
\begin{equation}
    p_j = \frac{s_j}{S}
\end{equation}
for \( i = 1, 2, \ldots, n - 1\).
The vector of probabilities \( {p_i} \) derived from the initial vector \( {v_i} \) is:
\begin{equation}
    {p_i} = \left[ \frac{s_1}{S}, \frac{s_2}{S}, \ldots, \frac{s_{n-1}}{S} \right]
\end{equation}

Each \( s_j \) in the vector \( {v_i} \) can take values from
a set that defines all possible states for that component.
The system state probability vector is then defined by:

\begin{equation}
    \label{equ:system-probabilities}
    P = [p_1, p_2, \ldots, p_n]
\end{equation}

The vector $P$ represents the probability of a complete system state,
combining state probabilities of all components into a single vector.
The component state reference probabilities \( r_i \) are constructed as
a uniform distribution.

\begin{equation}
    r_i = [u, u, \ldots, u]_{n-1}
\end{equation}

Where:
\( {r_i} \) is the reference state probability \(i\)-th component,
\( u \) represents uniform probability distribution calculated as \( {\frac{1}{n-1}}\)
The system state reference probability vector is then defined by:

\begin{equation}
    \label{equ:reference-probabilities}
    R = [r_1, r_2, \ldots, r_n]
\end{equation}

This vector $R$ represents the reference probability of the complete system state,
combining reference state probabilities of all components into a single vector.
An example Python implementation is provided in~\cite{Zizka2024calcscm}.

\section{Validation of SCM calculation for software systems}
\label{sec:validation}

Our goal is to validate that the SCM can be calculated for software systems
and that it follows the proposed theory. To have a controlled environment
where experiments can be executed, we need to define a software system model, which
may be executed using simulation software. The simulation results will provide
data for validation.

\subsection{Software system model design}
\label{subsec:model-design}

A system is a set of interconnected and interdependent
components forming an integrated whole. A software system is a system
composed of software components~\cite{myers2003software, valverde2003hierarchical}.
A software system can be
modeled as a network with nodes and connections between nodes. The topology
of the system is defined by connections among the nodes, some typical
topologies are~\cite{tanenbaum2003computer}:
\textbf{Full Mesh},
\textbf{Layered},
\textbf{P2P},
\textbf{Tree},
\textbf{Ring},
\textbf{Star},
\textbf{Bus},
\textbf{Hybrid}.

\begin{figure}[h]
    \centerline{\includegraphics[width=0.70\textwidth]{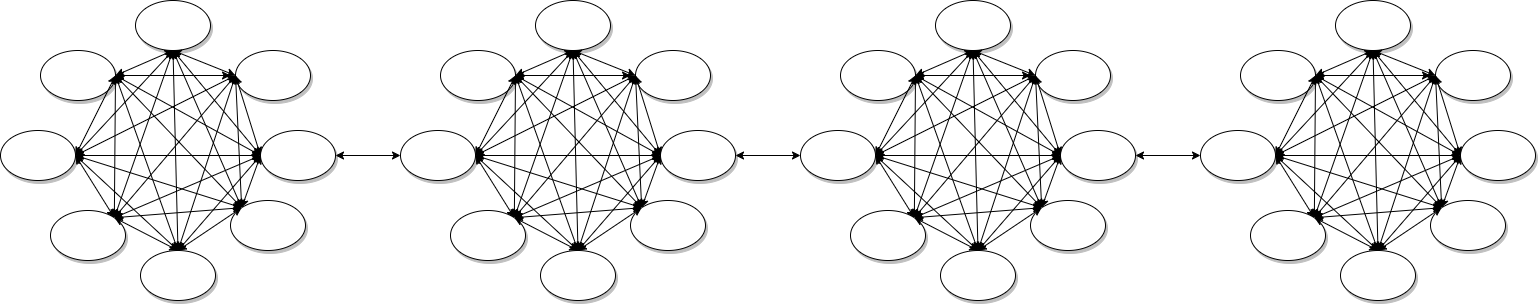}}
    \caption{Layered system}
    \label{fig:layered-system}
\end{figure}

Implementing each topology using the \textbf{Full Mesh} model
with configurable properties of the connections between nodes is possible. Such a model allows
flexibility when simulating different types of software system topologies.
The model complexity grows quadratically
and is \emph{$O(N^2)$}. For evaluating the statistical complexity,
the \textbf{Full Mesh} has an advantage as the probabilities for the
system can be determined directly, and simulations with 1024 nodes can still be
executed with reasonable time and resources. Considering the flexibility and
simplicity of calculations, we have chosen to use the \textbf{Full Mesh} model.
An example of a model diagram is in Figure~\ref{fig:layered-system}.
For purposes of verification of SCM, we’ll
be using \textbf{P2P}, \textbf{Full Mesh}, and \textbf{Layered} topologies,
which may represent ordered, chaotic, and structured types of systems, respectively.

Each system component can send or receive messages to or from
any other component. The states of the system are represented by
the states of each component, forming a vector defined by the amount
of communication with other components. By default, each component
will send or forward the received message to a random component generated
from a uniform probability distribution. This emulates chaotic, unstructured
communication and allows the creation of the \textbf{Chaotic} system
topology.

The model implementation allows the system to be partitioned into groups
of components by defining several groups. The components within the
group send and forward messages to randomly selected components
from a uniform probability distribution, similarly as in the case of
chaotic system, but only within the group of components. 
The groups of components represent layers in a multi-layer
network of components.
The model is designed so that if more than one
layer is configured, then 1\% of components on the layer send
and forward messages to a specific component in
another group in a strictly orderly configuration. This creates
a \textbf{Layered} topology with a system containing a different number
of layers.
To create a \textbf{P2P} topology, the model has the option to send
messages only to one component in the system or to a specific group of
components.
For purposes of verification of statistical complexity measure
calculation, we will create the following system models:

\textbf{Ordered} system model is designed to represent a fully
ordered system where only one state takes place out of all possible states.
The probability of this state is equal to one, and the probability
of any other possible state is zero. The expected entropy of
such a system is zero, and the disequilibrium is maximum for such
a system configuration. The resulting statistical complexity
is, therefore, zero. This model is realized by
a full mesh generic model where only one component sends
messages to another specific component. Only one state among
all possible states in the full mesh system is visited.

\textbf{Chaotic}
system model represents a system on the other
edge of the spectrum, where any state has an equal probability of occurrence.
This model is implemented by \textbf{Full Mesh} network where
every component sends a message to any component at random with
uniform probability distribution.

\textbf{Layered} system model is executed by partitioning
the system into 2, 4, 8, 16, 32, and 64 layers. The number of
layers is intentionally configured as powers of 2. Such systems’ resulting entropy
is also expected to be on the exponential power scale of 2 between ordered and
chaotic systems. This assumption defines the numerical
value of system configuration (see Table~\ref{tab:x-axis}).

\subsection{Simulation execution}
\label{subsec:simulation}

The model and system configurations were implemented using \emph{OMNeT++}.
The system component was implemented as a C++ \emph{Component} class
derived from \emph{cSimpleModule}. The system was modeled as
a network constructed as a full mesh using \emph{OMNeT++} modeling
language NED. The definition also contains configurable parameters
of the model for specifying the number of components, the number of
component groups (layers), and the definition of whether the component should
use random, uniformly distributed message targets or strictly predicted destinations.
The used model implementation is provided in \cite{Zizka2024model}.
The simulation was run with a different number of components
and different simulation times for the following set of system configurations:
\textbf{Ordered}, \textbf{Layered} with 64, 32, 16, 8, 4, and 2 layers and \textbf{Chaotic}.
The configurations are in detail described in Section~\ref{subsec:model-design}.
The simulation was run for systems with the following number of components:
128, 256, 512, and 1024. Simulation times were: 10, 100, 200, and 500 seconds.
The simulation times were selected based on experiments; longer simulation times
did not yield significant differences compared to results with a simulation time
of 500 seconds.
As highlighted by Feldman~\cite{Feldman1998}, defining the complexity’s
relationship to the system’s actual structure is essential.
As the models
have an exponentially increasing number of layers, we expect 
the system configuration dimension to be exponential.
We define the completely ordered system as having
dimensionality 1. The dimensionality of the configurations
is summarized in Table~\ref{tab:x-axis}.

\begin{table}[h]
    \centering
    \caption{System configuration dimensionality}
    \label{tab:x-axis}
    \begin{tabular*}{\hsize}{@{\extracolsep{\fill}}ccc@{}}
        \toprule
        Configuration & Dimension & Abbreviation \\
        \colrule
        Ordered    & 1 ($2^0$) & Or  \\
        64 Layers    & 2 ($2^1$) & L64  \\
        32 Layers    & 4 ($2^2$) & L32  \\
        16 Layers    & 8 ($2^3$) & L16  \\
        8 Layers    & 16 ($2^4$) & L8  \\
        4 Layers    & 32 ($2^5$) & L4  \\
        2 Layers    & 64 ($2^6$) & L2  \\
        Chaotic    & 128 ($2^7$) & Ch  \\
        \botrule
    \end{tabular*}
\end{table}

\subsection{Results}
\label{sec:results}

\begin{figure}[t]\vspace*{4pt}
    \centering
    \begin{subfigure}[b]{0.3\textwidth}
        \includegraphics[width=\textwidth]{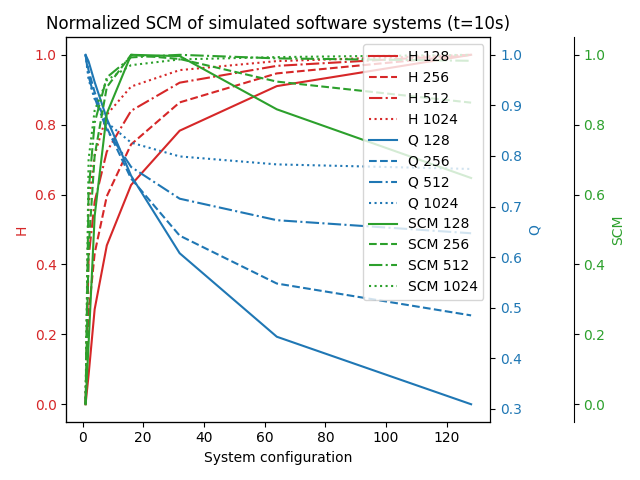}
    \end{subfigure}
    \begin{subfigure}[b]{0.3\textwidth}
        \includegraphics[width=\textwidth]{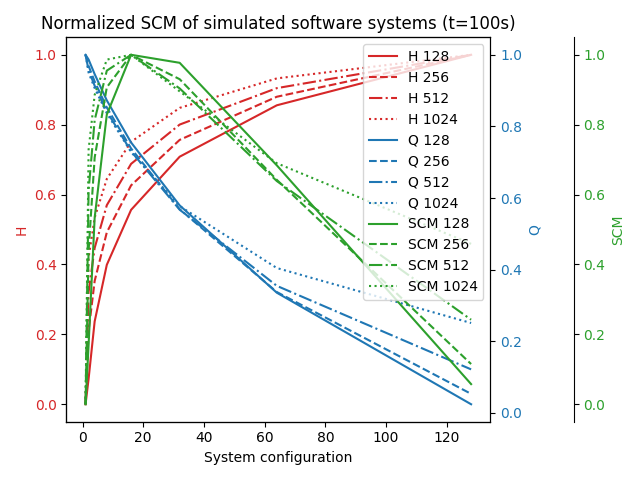}
    \end{subfigure}
    \begin{subfigure}[b]{0.3\textwidth}
        \includegraphics[width=\textwidth]{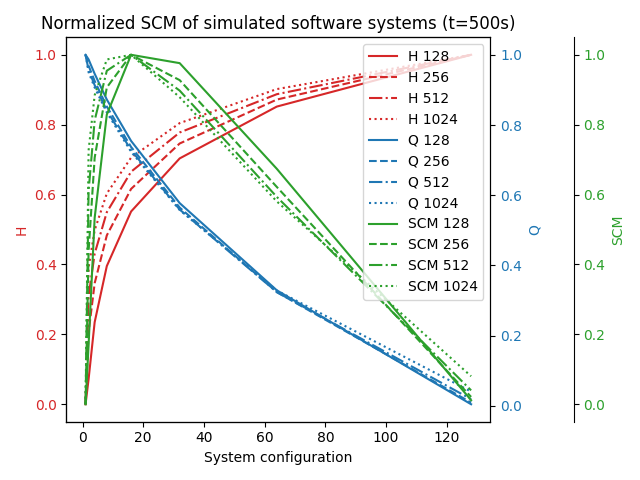}
    \end{subfigure}
    \caption{Normalized statistical complexity of simulated systems}
    \label{fig:normalized-simulated}
\end{figure}

The results of the simulations and calculation of normalized statistical complexity
are shown in Figure~\ref{fig:normalized-simulated}.
The data points for a simulation time of 500 seconds are shared in Table~\ref{tab:norm-results-500s}.
Complete results, including graphs and data tables, are provided in~\cite{Zizka2024data}.
The graph’s x-axis represents the system configuration or system
dimensionality described in Table~\ref{tab:x-axis},
ranging between ordered and chaotic systems. The y-axis provides
calculated entropy values $H$, disequilibrium $Q$, and
resulting statistical complexity measure $SCM$. The graphs
show simulations of different sizes of simulated systems
with 128, 256, 512, and 1024 components respectively coded in the
graphs' legend as $SCM n$ where $n$ represents the number of components.
Each graph shows results for different simulation times denoted
by $t=[time in seconds]$ for simulation times 10, 100, 200 and 500
seconds respectively.
Similar to the graphs, the Table~\ref{tab:norm-results-500s} shows calculated results for
each system configuration. The columns provide compound values of
$Q$, $H$ and $SCM$ for each system size $n$. The rows represent
the system configuration defined by Table~\ref{tab:x-axis}.
The normalized results of calculated SCM for simulated software systems,
as presented in Figure~\ref{fig:normalized-simulated} and in 
Table~\ref{tab:norm-results-500s}
show agreement with the theoretical
intuitive behavior of SCM as described in Figure~\ref{fig:scm-scatch}. 
The normalization allows for the comparison of different sizes
of simulated software systems, unlike the actual values where
magnitudes of SCM differ as expected~\cite{Zizka2024data}.

The results demonstrate that the absolute value of entropy $H$ for layered
configurations increases as the number of components increases. This
is expected as the system with more components contains more information.
The disequilibrium $Q$ converges to zero irrespective of the number of components
where the system configuration reaches a chaotic type of system.
This behavior is also expected as this is the measure of the organization of
a system. For an unordered system, the value should be low, irrespective
of the number of components of the system.
We can determine
the system configuration with the highest complexity from the obtained results by locating
a row with $\overline{SCM}$ equal to value $1.000$. For example, in
Table~\ref{tab:norm-results-500s}, the highest complexity was measured
for system L8. The system with eight layers and assigned dimensionality
16 based on Table~\ref{tab:x-axis}.
As the simulation time increases, the values of $Q$, $H$, and $SCM$
converge. With short simulation times, the amount of accumulated
information is low. Since the calculations are based on probabilities
of system state, the precision of the measurements depends
on the sample size.

The simulated software system model was designed to scale with
the number of components, and the maximum measured complexity
peaks with the same system clustering configuration. The graphs show
that with a higher amount of components, the complexity measurement
shifts towards a lesser number of layers. This shift results from
the fact that layers contain more components,
resulting in an overall higher amount of information. The same is
also visible in entropy values $H$, an expected consequence
of $SCM$ being a measure of the information in the system.
The complexity of the ordered system is low, as well
as the complexity of the disordered system. The system’s organization
into layers increases the measured complexity of the system, and
for the selected system design, the highest complexity is achieved
for a system with eight layers.

\begin{table}[h]
    \caption{Normalized simulation results for t=500s}
    \label{tab:norm-results-500s}
    \begin{tabular*}{\hsize}{@{\extracolsep{\fill}}crrrrrrrrrrrrr@{}}
        \toprule
        \multicolumn{1}{c}{} & \multicolumn{3}{c}{128} & \multicolumn{3}{c}{256} & \multicolumn{3}{c}{512} & \multicolumn{3}{c}{1024} \\
        \multicolumn{1}{c}{} & \multicolumn{1}{c}{$\overline{H}$} & \multicolumn{1}{c}{$\overline{Q}$} & \multicolumn{1}{c}{$\overline{SCM}$} & \multicolumn{1}{c}{$\overline{H}$} & \multicolumn{1}{c}{$\overline{Q}$} & \multicolumn{1}{c}{$\overline{SCM}$} & \multicolumn{1}{c}{$\overline{H}$} & \multicolumn{1}{c}{$\overline{Q}$} & \multicolumn{1}{c}{$\overline{SCM}$} & \multicolumn{1}{c}{$\overline{H}$} & \multicolumn{1}{c}{$\overline{Q}$} & \multicolumn{1}{c}{$\overline{SCM}$} \\
        \colrule
        Or & 0.000 & 1.000 & 0.000 & 0.000 & 1.000 & 0.000 & 0.000 & 1.000 & 0.000 & 0.000 & 1.000 & 0.000 \\
        L64 & 0.072 & 0.987 & 0.170 & 0.206 & 0.969 & 0.439 & 0.309 & 0.957 & 0.610 & 0.391 & 0.950 & 0.727 \\
        L32 & 0.235 & 0.946 & 0.534 & 0.346 & 0.925 & 0.705 & 0.431 & 0.913 & 0.812 & 0.498 & 0.907 & 0.884 \\
        L16 & 0.395 & 0.873 & 0.829 & 0.483 & 0.853 & 0.907 & 0.550 & 0.842 & 0.954 & 0.603 & 0.836 & 0.987 \\
        L8 & 0.552 & 0.755 & 1.000 & 0.616 & 0.738 & 1.000 & 0.665 & 0.728 & 1.000 & 0.706 & 0.724 & 1.000 \\
        L4 & 0.703 & 0.578 & 0.976 & 0.746 & 0.565 & 0.928 & 0.778 & 0.559 & 0.898 & 0.805 & 0.558 & 0.879 \\
        L2 & 0.852 & 0.329 & 0.674 & 0.872 & 0.324 & 0.622 & 0.887 & 0.324 & 0.593 & 0.902 & 0.329 & 0.580 \\
        Ch & 1.000 & 0.005 & 0.012 & 1.000 & 0.010 & 0.021 & 1.000 & 0.019 & 0.040 & 1.000 & 0.041 & 0.080 \\
        \botrule
    \end{tabular*}
\end{table}

\section{Limitations} \label{sec:limitations}

Although it was shown that the statistical complexity measure
can be calculated for software systems modeled as multi-layer
networks, the method comes with limitations that must be
considered.
As commented by Feldman~\cite{Feldman1998}, the statistical complexity
requires an understanding of the interpretation of the system’s
structure calculated using disequilibrium $Q$. The
measured values must have concrete representations. This might
be challenging for software systems
viewed as a black box. 
Knowing the software system design will allow for unambiguous
interpretation, as
demonstrated by the software system design used
in our research.
The definition of system state vector by Equation~\ref{equ:system-probabilities}
and selection of reference system probabilities defined by Equation~\ref{equ:reference-probabilities}
is critical. It might not be easy to derive these for all
software systems, and universally usable and validated methods
must be developed.
Understanding that statistical complexity
gives just one specific view of a software system is essential. 
Other methods have to be utilized to interpret the measured data,
such as qualitative analysis of the system
properties and comparison of estimated statistical
complexity between real software systems and system models.
Digital twin technologies could be utilized.
The statistical complexity is based on system state
probabilities, so the measurement duration must be defined carefully.

\section{Conclusions} \label{sec:conclusions}

We have demonstrated that statistical complexity measure (SCM)
can be utilized to measure the complexity of a software system modeled as a
network composed of components and interconnections.
The measured complexity is reflecting
both the amount of information in the system as well as the
structure or organization of the system. We have also shown
how the state vector of such a system can be designed in order
to collect probabilities of the system state necessary for the
SCM calculation. The hypothesis about complexity measure
values as presented by López-Ruiz~\cite{LopezRui1995} were validated
using simulation of a software system with varying organization
structure emulating a system as a multi-layer network.
The SCM measurements on simulated systems show good agreement
with the theory.
We have demonstrated the importance of normalization of the
calculated values, allowing comparison of software systems
of different sizes. This has direct implications, showing
that the SCM may be used to compare systems
based on complexity measures.
The complexity of a system can be approached from different angles,
a complex software system can be defined by a set of properties~\cite{zizka2023towards},
however, such a definition does not
provide a means to measure and compare the systems quantitatively.
The statistical complexity measure can be used as one of
the perspectives describing complex software systems, providing
a qualitative measure.
The qualitative measure can then be used to compare
software systems based on the SCM and analyze
software systems by comparing them to models. This measure may then reveal
relationships between software system organization
and information contained in the system, and such knowledge may be utilized to
understand the effects of complex software system properties as
presented in~\cite{zizka2023towards}. 
There is a potential to use SCM
to assist in the optimization of software systems to maximize the
system complexness providing emergent behaviors, such optimization may be implemented as an adaptive
self-organization of the software systems based on SCM
feedback values. The measured SCM may also gain insight into
unwanted emergent behaviors, such as emergent errors,
increased security attack space and the system being complicated
rather than complex, further increasing maintenance and
development costs.
We will, in future research, focus on these aspects and on
utilization of SCM in real software systems bridging results
provided by this paper through simulation into practice.
These possibilities open new directions
of research toward understanding complex software systems.

\section*{Acknowledgements}

The author would like to express his profound gratitude to Dr. Bruno Rossi and
Prof. Tomáš Pitner for their invaluable comments and thorough review, which enhanced
the quality and clarity of the text of this paper.

\bibliography{ref}
\bibliographystyle{elsarticle-harv}

\end{document}